\documentclass[12pt]{article}

\usepackage{graphicx}
\usepackage{multicol} 
\usepackage{color}

\usepackage{amsmath,amssymb,mathrsfs}

\definecolor{rosso}{cmyk}{0,1,1,0.4}
\definecolor{rossos}{cmyk}{0,1,1,0.55}
\definecolor{rossoc}{cmyk}{0,0.5,1,0.2}
\definecolor{blu}{cmyk}{1,1,0,0.3}
\definecolor{blus}{cmyk}{1,1,0,0.6}
\definecolor{blucc}{cmyk}{1,0.4,0.2,0}
\definecolor{viola}{cmyk}{0,1,0,0.6}
\definecolor{viola2}{cmyk}{0,1,0.2,0.6}
\definecolor{verde}{cmyk}{0.92,0,0.59,0.25}
\definecolor{verdec}{cmyk}{0.92,0,0.59,0.15}
\definecolor{verdes}{cmyk}{0.92,0,0.59,0.4}
\font\tenrsfs=rsfs10 at 12pt
\font\sevenrsfs=rsfs7
\font\fiversfs=rsfs5
\newfam\rsfsfam

\textfont\rsfsfam=\tenrsfs
\scriptfont\rsfsfam=\sevenrsfs
\scriptscriptfont\rsfsfam=\fiversfs
\def\mathscr#1{{\fam\rsfsfam\relax#1}}

\newcommand{\SU}{{\rm SU}}

\newcommand{\fig}[1]{~\ref{fig:#1}}
\newcommand{\eq}[1]{~{\rm (\ref{eq:#1})}}

\newcommand{\GeV}{\,{\rm GeV}}
\newcommand{\TeV}{\,{\rm TeV}}
\def\circa#1{\,\raise.3ex\hbox{$#1$\kern-.75em\lower1ex\hbox{$\sim$}}\,}

\newcommand{\NP}{Nucl. Phys.}

\newcommand{\PR}{Phys. Rev.}
\newcommand{\beq}{\begin{equation}}
\newcommand{\eeq}{\end{equation}}
\newcommand{\bea}{\begin{eqnarray}}
\newcommand{\eea}{\end{eqnarray}}

\def\circa#1{\,\raise.3ex\hbox{$#1$\kern-.75em\lower1ex\hbox{$\sim$}}\,}
\makeatletter

\def\art{\@ifnextchar[{\eart}{\oart}}
\def\eart[#1]#2#3#4#5#6{{\rm #2}, {\em #3 \rm #4} {\rm (#6) #5} [{#1}]}
\def\hepart[#1]#2{{\rm #2, #1}}
\newcommand{\oart}[5]{{\rm #1}, {\em #2 \rm #3} {\rm (#5) #4}}
\newcommand{\y}{{\rm and} }
\newcounter{alphaequation}[equation]
\def\thealphaequation{\theequation\hbox to
0.6em{\hfil\alph{alphaequation}\hfil}}
\def\eqnsystem#1{
\def\@eqnnum{{\rm (\thealphaequation)}}
\def\@@eqncr{\let\@tempa\relax \ifcase\@eqcnt \def\@tempa{& & &} \or
  \def\@tempa{& &}\or \def\@tempa{&}\fi\@tempa
  \if@eqnsw\@eqnnum\refstepcounter{alphaequation}\fi
\global\@eqnswtrue\global\@eqcnt=0\cr}
\refstepcounter{equation} \let\@currentlabel\theequation \def\@tempb{#1}
\ifx\@tempb\empty\else\label{#1}\fi
\refstepcounter{alphaequation}
\let\@currentlabel\thealphaequation
\global\@eqnswtrue\global\@eqcnt=0 \tabskip\@centering\let\\=\@eqncr
$$\halign to \displaywidth\bgroup \@eqnsel\hskip\@centering
$\displaystyle\tabskip\z@{##}$&\global\@eqcnt\@ne
\hskip2\arraycolsep\hfil${##}$\hfil& \global\@eqcnt\tw@\hskip2\arraycolsep
$\displaystyle\tabskip\z@{##}$\hfil
\tabskip\@centering&\llap{##}\tabskip\z@\cr}
\def\endeqnsystem{\@@eqncr\egroup$$\global\@ignoretrue} \makeatother

\newcommand{\sW}{s_{\rm W}}
\newcommand{\cW}{c_{\rm W}}

\begin{document}

\thispagestyle{empty}

\begin{flushright}
{
KA-TP-02-2005\\
IFUP--TH/2005-02\\
hep-ph/0502095\\

}
\end{flushright}
\vspace{1cm}

\begin{center}
{\LARGE \bf \color{rossos}
Supersymmetry and precision data after LEP2}\\[1cm]

{
{\large\bf Guido Marandella$^a$,  Christian Schappacher}$^b$,
{\large\bf Alessandro Strumia}$^{c}$
}  
\\[7mm]
{\it $^a$ Theoretical Division T-8, Los Alamos National Laboratory, Los Alamos, NM 87545, USA}\\[3mm]
 {\it $^b$ Institut f{\"u}r Theoretische Physik, Universit{\"a}t Karlsruhe, Germany}\\[3mm]
{\it $^c$ Dipartimento di Fisica dell'Universit{\`a} di Pisa and INFN, Italia}\\[1cm]
\vspace{1cm}
{\large\bf\color{blus} Abstract}

\end{center}
\begin{quote}
{\large\noindent\color{blus}

We study one loop supersymmetric corrections to precision observables.
Adding LEP2 $e\bar{e}\to f\bar{f}$ cross sections to the data-set removes
previous hints for SUSY and the resulting constraints are in some cases
stronger than direct bounds on sparticle masses.
We consider specific models: split SUSY,
CMSSM, gauge mediation, anomaly and radion mediation.
Beyond performing a complete one-loop analysis, we also develop
a simple approximation, based on the $\hat S,\hat T,W,Y$ `universal' parameters.
SUSY corrections give $W,Y>0$ and mainly depend on the left-handed slepton and squark masses,
on $M_2$ and on $\mu$.
}
\end{quote}

\setcounter{page}{1}
\setcounter{footnote}{0}

\section{Introduction}
In spite of the competition from alternative proposals, low-energy supersymmetry (SUSY)
remains the most promising interpretation of
the origin of the  electroweak symmetry breaking scale.
One important virtue of supersymmetry is that (after imposing matter parity) precision observables
do not receive tree-level corrections.
Therefore direct searches, limited by the collider energy,
provide the dominant constraints.
In this paper we study supersymmetric one-loop corrections to precision observables,
which allow indirect tests and constraints.

We add to the data-set the LEP2  $e\bar{e}\to e\bar{e},\mu\bar\mu,\tau\bar\tau,q\bar{q}$ 
cross sections,
not included in previous analyses
(see~\cite{Cho,A...,MTW} for some recent works).
A simple estimate shows that these LEP2 precision data have an important impact.
LEP2 observed $N\approx 10^4$ $e\bar{e}\to f\bar{f}$ events at center-of-mass energy
$\sqrt{s}\approx 200\GeV$.
Therefore LEP2 is sensitive to four-fermion operators, normalized as 
 $$\frac{4\pi}{\Lambda^2} (\bar{e}\gamma_\mu e)(\bar f \gamma_\mu f),\qquad\hbox{
 up to }\qquad \Lambda \approx \sqrt{\frac{sN^{1/2}}{\alpha}}\approx 10\TeV.$$
 Indeed LEP2 collaborations claim $\Lambda\circa{>}  10\TeV$~\cite{LEPEWWG,LEP2}.
 Supersymmetric particles of mass $m_{\rm SUSY}$  generate such operators at one-loop with coefficients
 $4\pi/\Lambda^2 \sim {g^4}/{(4\pi m_{\rm SUSY})^2} $.
This means
 $m_{\rm SUSY}\circa{>} g^2\Lambda/(4\pi)^{3/2}\approx 100\GeV$,
 which is comparable to direct collider bounds.\footnote{As an aside remark,
 we point out the relevance of LEP2 in a different context.
By performing a SM fit, we find that LEP2 data determine the weak angle $\sW\equiv \sin\theta_{\rm W}$
as accurately as low-energy measurements of $\sW$
i.e.\  5 times less accurately than $Z$-pole data.
Therefore future reports of $\sW$ measurements at different energies
should take LEP2 into account.}
This estimate motivates the present work, where we perform a full one-loop analysis (i.e.\ we include all one-loop 
 propagator, vertex and box diagrams) of LEP2 and traditional precision data.
 
We also develop a simple understanding,
based on the `heavy universal' approximation,
giving explicit analytical approximation for the
supersymmetric corrections to the $\hat S,\hat T,W,Y$ parameters~\cite{Barbieri:2004qk}.
This approximation is correct within $\sim30\%$ accuracy
and becomes exact in various limits.

The paper is organized as follows.
In section~\ref{STWY} we  give results for $\hat{S},\hat{T},W,Y$
and compare general features of corrections to precision data
to other indirect probes.
In section~\ref{split} we consider `split' supersymmetry~\cite{anthropoids}, which is a
simple warming exercise towards a full analysis.
In section~\ref{LIGHT} we consider a model with a simple spectrum.
In section~\ref{CMSSM} we consider the MSSM with unified soft terms at the GUT scale.
In section~\ref{GM} we consider gauge mediation models~\cite{GM}.
In section~\ref{AM} we study anomaly plus radion mediation~\cite{AM,RSS}.
In section~\ref{end} we conclude and summarize our results.

\section{Supersymmetric effects in $\hat{S},\hat{T},W,Y$ approximation}\label{STWY}
After imposing matter parity, supersymmetric corrections to precision
observables arise dominantly at one loop.
Results greatly simplify in `heavy universal' approximation:
i.e.\ one assumes that new physics is above the weak scale (`heavy'),
and that couples dominantly to vector bosons (`universal').
At first sight the `heavy universal' 
approximation is not applicable to the case of supersymmetry
because:
\begin{enumerate}
\item SUSY is not `universal':
corrections to gauge boson propagators, to vertices and box diagrams
give comparable effects.
\item SUSY  is not `heavy': sparticle masses are expected to be comparable to the $Z$ mass.
\end{enumerate}
Actually the `heavy universal' approximation, in which all effects can be encoded in the four
 $\hat{S},\hat{T},W,Y$  parameters~\cite{Barbieri:2004qk}, is useful because:
\begin{enumerate}
\item Gauginos, higgsinos, Higgs bosons alone (with heavy sfermions)
are universal because negligibly couple to light fermions.
Sfermions alone (with heavy gauginos, higgsinos, Higgs bosons) are also universal.
In the most natural case where all sparticles are light,
vertex and box diagrams give non-universal effects.
However universal corrections
are cumulative in the number of generations (and of colors and of weak components), 
while non-universal corrections are not.
Therefore the universal approximation is expected to hold within $1/N_{\rm gen}\sim 30\%$ accuracy,
and to become exact in various limits.
In universal approximation, all $Z$-pole observables and the $W$ mass 
can be condensed
in three numbers: $\varepsilon_1,\varepsilon_2,\varepsilon_3$~\cite{ABC}, 
which have been frequently employed in MSSM analyses.

\item Constraints from direct searches force almost all sparticles to be heavier than $m_{\rm SUSY}\circa{>}E_{\rm LEP2}$ where
$E_{\rm LEP2}\approx 100\GeV$ is the beam energy of LEP2.
Indeed the limiting factor at LEP2 was kinematics, rather than
production cross-sections and luminosity.
The kinematical threshold in one loop diagrams with two sparticles is $2m_{\rm SUSY}$, so that 
the simple `heavy sparticles' limit approximates SUSY corrections $\delta\varepsilon_{1,2,3}$ to
$Z$-pole observables within $(M_Z/2m_{\rm SUSY})^2< 25\%$ accuracy
as
\beq\label{sys:eps123}
 \delta\varepsilon_1 \simeq  \widehat{T }-W  - Y \frac{\sW^2}{\cW^2}\, , \qquad
\delta\varepsilon_2 \simeq  -W\, ,\qquad
\delta\varepsilon_3 \simeq  \widehat{S} -W-Y\, .
\eeq
Similarly, low-energy observables depend on other combinations of $\hat{S},\hat{T},W,Y$~\cite{Barbieri:2004qk}.
At LEP2 the `heavy' approximation fails by $(E_{\rm LEP2}/m_{\rm SUSY})^2\sim 1$, but remains
qualitatively correct. Indeed the missed effect is the resonant enhancement
of virtual effects present when fermionic sparticles are just above the LEP2 kinematical limit.
\footnote{Virtual effects of scalar particles are not resonantly enhanced,
because their coupling to a vector vanish in the non-relativistic limit.
As well known, their direct production is suppressed by the same factor:
as demanded by conservation of angular momentum
non-relativistic scalars cannot be produced in $s$-wave.}

\end{enumerate}
\begin{figure}
\begin{center}
$$\includegraphics[width=17.6cm]{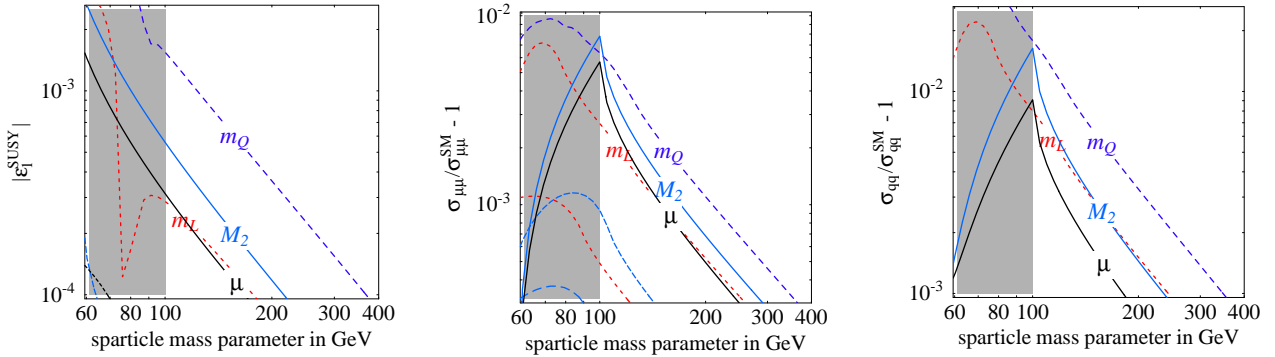}$$
\caption{\em Corrections to $\varepsilon_1$ and to $\sigma(e\bar{e}\to \mu\bar\mu,\sum_q q\bar{q})$
at CM energy $200\GeV$ generated when
only the sparticle with the indicated mass is light.
This allows to see the accuracy of the $\hat{S},\hat{T},W,Y$ approximation,
which gives straight lines with the correct asymptotic behavior for $m_{\rm SUSY}\gg M_Z$.\label{fig:single}}
\end{center}
\end{figure}
This feature is illustrated in fig.\fig{single}, where we show
full numerical results for the corrections to a few observables.\footnote{The SM
value of LEP2 observables is defined using the observables
$\alpha,M_Z,M_W$ to fix the SM parameters $v,g,g'$.
Alternative choices, such as $\alpha,M_Z,G_{\rm F}$, would give a different result
as discussed in appendix~\ref{tech}.}
We consider a variety of sparticle spectra where only one
kind of sparticles is light: only left-handed sleptons, left-handed squarks, only gauginos,
only Higgsinos, and so on.
Their masses are respectively determined by the parameters $m_{\rm SUSY}=\{m_L,m_Q,\mu,M_2,\ldots\}$.
In these limits the approximation 1 (`universal' SUSY) becomes exact. 
Therefore, by making also approximation 2 (`heavy' SUSY),
gives the correct asymptotic $1/m_{\rm SUSY}^2$ behavior
for $m_{\rm SUSY}\gg M_Z$, which would 
correspond to straight lines in fig.\fig{single}.
We see that all curves remain roughly straight in all the allowed range $m_{\rm SUSY}\circa{>}100\GeV$.
Approximation 2 badly fails only for $m_{\rm SUSY} \circa{<}100\GeV$, which is now excluded.

Later we give some examples of the accuracy of our approximations.
We skip a detailed discussion of the accuracy of approximation~1 because
it is not new, and because there is no simple general way of comparing approximation~1 with full results.  We remark that since the two approximations introduce comparable errors,
to really improve the accuracy one needs to avoid both approximations.
E.g.\ analyses performed dropping approximation 2 (`heavy' SUSY)
but making the approximation 1 (`universal' SUSY i.e.\ corrections to vertices
and boxes are neglected) are much more complicated than our approximate analysis,
without being significantly more accurate.
 
 \medskip
 
 We also perform a full one-loop analysis,
 supplementing the traditional precision data\footnote{
 We do not perform a full analysis of  low energy observables
 (atomic parity violation, M\o{}ller scattering, neutrino/nucleon scattering)
 because they have a minor impact in the global fit
 and each of them would require a dedicated computation.
 Corrections to low energy observables are included in $\hat{S},\hat{T},W,Y$ approximation.}
 with LEP2 $e^+ e^-\to  e^+e^-,\mu^+\mu^- ,\tau^+ \tau^-,q\bar{q}$ cross sections.
Almost all needed pieces of the computations can be obtained from literature:
SUSY corrections
to gauge boson propagators can be found in~\cite{MSSM1, Matchev96,MSSMPi,Cho},
to $Z$-boson vertices in~\cite{MSSMZvertex,Cho},
to $\mu$-decay in~\cite{MSSM1,MSSMmudecay2,Cho,Matchev96},
to LEP2 $e\bar{e}\to f\bar{f}$  cross sections  in~\cite{MSSMLEP2}
(only for $f\neq e$).
We have recomputed all LEP2 cross sections, including for the first time $e^+ e^-\to  e^+e^-$,
using the FeynArts, FormCalc and LoopTools codes~\cite{FormCalc,LoopTools}.
Technical details are given in appendix~\ref{tech}.

\begin{figure}
\begin{center}
$$\begin{array}{cccc}\displaystyle
{\color{blus}\widehat{S}} = \frac{g}{g'}\Pi'_{W_3 Y}(0) & \displaystyle
{\color{blus} \widehat{T} }= \frac{\Pi_{W_3 W_3}(0)-\Pi_{W^+W^-}(0)}{M_W^2}& \displaystyle
\quad{\color{blus} W} = \frac{M_W^2}{2}\Pi''_{W_3 W_3}(0) \quad& \displaystyle
{\color{blus} Y} =\frac{M_W^2}{2}\Pi''_{YY}(0) \\ \displaystyle
\quad(H^\dagger \tau^a H) W^a_{\mu\nu} Y_{\mu\nu}\quad & \displaystyle
|H^\dagger D_\mu H|^2& \displaystyle
\frac{(D_\rho W^a_{\mu\nu})^2}{2}& \displaystyle
\frac{(\partial_\rho Y_{\mu\nu})^2}{2}\end{array}$$
\includegraphics[width=16cm]{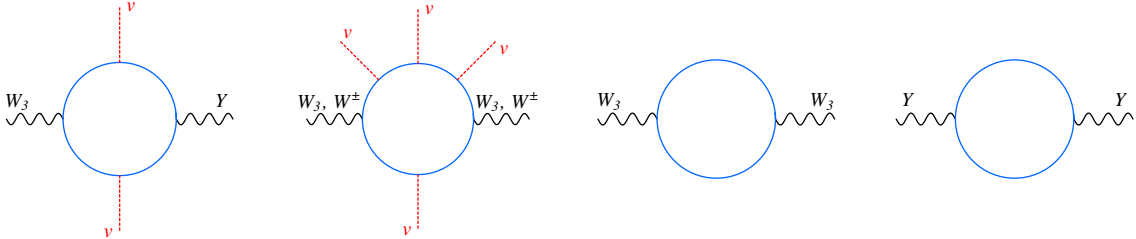}
\caption{\em Upper row: definition of $\hat S$, $\hat T$, $W$ and $Y$
in terms of canonically normalized inverse propagators $\Pi$.
Middle row: the corresponding dimension 6 operators.
Lower row: one-loop Feynman graphs that contribute to $\hat S$, $\hat T$, $W$ and $Y$.
Unspecified lines denote generic sparticles.
\label{fig:FeynSTWY}}
\end{center}
\end{figure}

We now give simple analytic expressions for SUSY effects in `heavy universal' approximation.
We list how each sparticle contributes to $\hat{S},\hat{T},W,Y$.
The contributions to $\hat{S},\hat{T},W,Y$ are obtained computing 
the diagrams in the lower row of fig.\fig{FeynSTWY}, 
that directly correspond to the dimension 6 operators listed in the middle row of fig.\fig{FeynSTWY}.
Sometimes our analytic approximations have a definite sign, 
confirming some results previously noticed by performing numerical scans.\footnote{
Virtual SUSY effects are present also in the QCD sector.
In `universal heavy' approximation all such QCD effects are encoded in the $Z$ parameter,
defined as $Z = M_W^2\Pi''_{GG}(0)/2$~\cite{Barbieri:2004qk} and given by
\beq Z = \frac{\alpha_3}{80\pi}M_W^2 \bigg[ \frac{2}{m_Q^2}+\frac{1}{m_U^2}+\frac{1}{m_D^2} +
\frac{8}{M_3^2}\bigg].\eeq
}

\subsection{Sfermions}\label{sec:sfermions}
The three generations of sfermions with masses $m_L,m_E,m_Q,m_U,m_D$ 
and hypercharges
\begin{equation}\label{eq:Y}
  Y_L=-1/2, \;\;\;\; 
  Y_E=1, \;\;\;\; 
  Y_Q=1/6, \;\;\;\; 
  Y_U=-2/3, \;\;\;\; 
  Y_D=1/3.
\end{equation}
give
\begin{eqnsystem}{sys:sfermions}
\hat{S}&=&- \frac{\alpha_2}{24\pi} \bigg[ M_W^2\bigg( 3\frac{Y_L}{m_L^2}+9\frac{Y_Q}{m_Q^2}\bigg)\cos2\beta+\frac{1}{2}
\frac{m_t^2}{m_Q^2}\bigg],
\\
\hat{T} &= &\frac{\alpha_2}{16\pi} M_W^2 \cos^22\beta\bigg( \frac{1}{m_L^2}+\frac{2}{m_Q^2}\bigg)
+\hat{T}_{\rm stop},\\
  Y&=&\frac{\alpha_Y}{40 \pi}  M_W^2 \bigg(\frac{Y_E^2}{m_E^2}+ 2 \frac{
  Y_L^2}{m_L^2}+ 3 \frac{Y_D^2}{m_D^2}+ 3 \frac{Y_U^2}{m_U^2} +  6
  \frac{Y_Q^2}{m_Q^2} \bigg),\\
  W&=&\frac{\alpha_2}{80 \pi}  M_W^2  \bigg(\frac{
  1}{m_L^2}+  
  \frac{3}{m_Q^2} \bigg).
\end{eqnsystem}
We assumed that sfermions within each generation have the same mass: $m_{L_1} = m_{L_2} = m_{L_3}\equiv m_L$ and so on.
If sfermions of different generations have instead different masses our 
above expressions can be immediately generalized.
The leading order approximation to the stop contribution,
$$
\hat{T}_{\rm stop}=\frac{\alpha_2}{16\pi}\frac{(m_t + M_W\cos2\beta)^2}{m_{Q_3}^2 M_W^2}$$
is often not accurate enough,
mainly because of the possibility of sizable stop mixing $\theta_{\tilde{t}}$
induced by $A_t$ is neglected.
This effect could be taken approximatively into account, 
but we prefer to use the exact expression for $\hat{T}_{\rm stop}$  $$\hat{T}_{\rm stop} = \frac{3\alpha_2}{16\pi M_W^2}\bigg[\cos^2\theta_{\tilde{t}}\, f(m^2_{\tilde{t}_1},
m^2_{\tilde{b}_L})+\sin^2\theta_{\tilde{t}}\,  f(m^2_{\tilde{t}_2},
m^2_{\tilde{b}_L})-\sin\theta_{\tilde{t}}\cos\theta_{\tilde{t}}\, f(m^2_{\tilde{t}_1},
m^2_{\tilde{t}_2})\bigg]$$
because it is so simple
that expanding it cannot give a significant simplification.
In the above equations $m_{\tilde{t}_{1,2}}$ are stop masses, 
$f(a,b) = a+b+2ab \ln(a/b)/(b-a)$.
For large $\tan\beta$ one needs to take into account also sbottom mixing.

Notice that $W,Y>0$.
Furthermore $\hat{S}$ is almost always negative
(the positive $Y_Q$ contribution is dominant only if 
squarks of the first two generations are enough lighter than sleptons and stops).
The factor $\cos 2\beta=(v_d^2-v_u^2)/v^2 < 0$ is associated to
custodial symmetry breaking.

\subsection{MSSM Higgs bosons}
Their properties are determined, at tree level,
by two parameters: $\tan\beta$ and the mass of the
pseudo-scalar Higgs $m_A = [m_{h_{\rm u}}^2+m_{h_{\rm d}}^2+2|\mu|^2]^{1/2}$.
We find
\begin{eqnsystem}{sys:higgs}
\hat{S}&=&-\frac{\alpha_2}{48\pi} \frac{M_W^2}{m_A^2}(1-\frac{M_Z^2}{2M_W^2}\sin^22\beta),\\
\hat{T} &= &\frac{\alpha_2}{48\pi} \frac{M_W^2}{m_A^2}(1-\frac{M_Z^2}{M_W^2}\sin^2 2\beta),\\
  Y&=&\frac{\alpha_Y}{240 \pi}    \frac{M_W^2}{m_A^2},\\
  W&=&\frac{\alpha_2}{240 \pi}    \frac{M_W^2}{m_A^2}.
  \end{eqnsystem}
The Higgs contributions to $W,Y$ are always positive,
the contribution to $\hat{S}$ is negative, 
the contribution to $\hat{T}$ is positive for $\tan\beta\circa{>} 1.7$.
Higgs bosons contribute to  $W,Y$ in the same way as one slepton doublet with $m_L \leftrightarrow m_A$.
Their contributions to $\hat{S},\hat{T}$ are instead different, because
$\SU(2)_L$-breaking affects differently Higgs masses and slepton masses.
MSSM Higgs bosons give `small' corrections to $\hat{S},\hat{T},W,Y$.

\subsection{Gauginos and higgsinos}
Their contributions to $\hat{S},\hat{T},W,Y$ are obtained computing 
the diagrams in fig.\fig{FeynSTWY}.
The resulting expressions in terms of $M_1,M_2,\mu,\beta$ are explicit
(unlike the usual ones, written as sums over chargino and neutralino eigenstates
appropriately weighted by their eigenvectors) but lengthy.\footnote{The full expression,
not reported here but available in Mathematica format,
is used in our later computations.}
Since we anyhow expect that for light sparticles
our approximations are accurate within $\sim 30\%$,
for simplicity we here report them neglecting terms suppressed by 
$s_{\rm W}^2 \approx 0.22$.
In this limit $\hat{S},\hat{T}$ no longer depend on $M_1$.
The terms that we neglect vanish in  the limit $M_1\gg M_2, |\mu|$.

No further approximation is needed for $W$ and $Y$, which are given by simple expressions.
Gauginos and higgsinos contribute as
\begin{eqnsystem}{sys:gahi}
\hat{S}&= &\frac{\alpha_2 M_W^2}{12\pi M_2^2}\bigg[ \frac{r(r-5-2r^2)}{(r-1)^4}+
\frac{1-2r+9r^2-4r^3+2r^4}{(r-1)^5}\ln r \bigg] +\\
&&+ \frac{\alpha_2 M_W^2 }{24\pi M_2\mu}\bigg[
 \frac{2-19r+20r^2-15r^3}{(r-1)^4}+\frac{2+3r-3r^2+4r^3}{(r-1)^5}2r\ln r\bigg]\sin 2\beta
\nonumber ,\\
\hat{T} &=&\frac{\alpha_2 M_W^2}{48\pi M_2^2}\bigg[ \frac{7r-29+16r^2}{(r-1)^3}+
\frac{1+6r-6r^2}{(r-1)^4}6\ln r\bigg]\cos^2 2\beta, \\
  Y&=& \frac{\alpha_Y}{30 \pi} \frac{M_W^2}{\mu^2},\\
  W&=& \frac{\alpha_2}{30 \pi} \bigg[ \frac{M_W^2}{\mu^2} +  \frac{2 M_W^2}{M_2^2} \bigg]\end{eqnsystem}
where $r = \mu^2/M_2^2$. 
While $W$ and $Y$ are positive, $\hat{S}$ and $\hat{T}$ can have both signs.
The second contribution to $\hat{S}$ is suppressed by one power of $\tan\beta$.
Notice that for $|\mu| \ll M_1,M_2$ or for $|\mu| \gg M_2,M_1$
$\hat{S}$ and $\hat{T}$ are negligible because suppressed by the heaviest mass,
while corrections to $W$ and $Y$ can still be sizable because
suppressed by the lightest mass.

\subsection{General features}
Presently SUSY models have many  unknown parameters and there is no experimental
evidence for SUSY:
a general analytic understanding seems more appropriate than
detailed numerical analyses focussed on particular points.

All contributions to $\hat S,\hat T,W,Y$ are of order 
$(gM_W/4\pi m_{\rm SUSY})^2 $,
with the exception of corrections due to stops.
As well known, and as clear from fig.\fig{FeynSTWY},
stops give corrections to $\hat{T}$ enhanced by $(\lambda_t/g)^4$
and corrections to $\hat{S}$ enhanced by $(\lambda_t/g)^2$.
This is the largest single effect if $m_{Q_3}$ is small enough.

Otherwise the total effect can still be detectable, in view
of the cumulative effect of the large number of sparticles.
E.g.\ all sparticles give positive contributions to $W,Y$.
Setting all sparticles to a common mass $m_{\rm SUSY}$ one gets
$W\approx  10^{-3}({100\GeV}/{m_{\rm SUSY}})^2$.
Without including LEP2, data are compatible with the SM with
a mild preference for a positive $W$ (i.e.\ a negative correction to $\varepsilon_2$).
This preference disappears when LEP2 is included
and the measured value of $W$ disfavors at almost $2\sigma$
having all these sparticles as light as allowed by direct experimental bounds,
$m_{\rm SUSY}\approx 100\GeV$.

\medskip


One important feature of the above expressions for $\hat{S},\hat{T},W,Y$ 
is the absence of notable features, 
either suppressions or enhancements.
Furthermore SUSY corrections to precision observables
mainly depend on a few SUSY parameters: 
left-handed slepton and squark masses, $M_2$ and $\mu$.
This makes precision data a `robust' generic test,
unlike other effects (like $b\to s\gamma$, $g_\mu-2$, $B_s \to \mu\bar\mu$)
 which can be strongly enhanced in some regions of the parameter space 
 (e.g.\ at large $\tan\beta$\footnote{Notice however that
large $\tan\beta$ is
 naturally obtained for small $\mu$ such that $\mu\tan\beta$ is not enhanced.
 Unlikely accidental cancellations are needed to get large $\tan\beta$ and large $\mu$.})
 and negligible in other regions (e.g.\ in the `split' SUSY limit~\cite{anthropoids}).
 Another example is  the predicted abundancy of thermal dark matter,
  which is suppressed in corners of the parameter space where
  the two lightest sparticles happen to be quasi-degenerate~\cite{coann}.

\bigskip

Furthermore corrections to precision observables are `robust' under variations of the SUSY model.  
E.g.\ the non-observation of the lightest Higgs might be interpreted as an effect of
some extra physics that increases the MSSM prediction for the Higgs mass.
The simplest option is adding a singlet chiral superfield
coupled to the Higgs doublets.
Such extra singlet can drastically affect also dark matter signals of SUSY,
but has almost no impact on precision tests.

We will consider several models, showing contours in the $(m_0, M_{1/2})$ plane (in the CMSSM,
and analogous ones in other models)
with the $\mu$ parameter determined from the condition of correct electroweak symmetry breaking.
This kind of plots hides the fact that
most of the SUSY parameter space stays at light $m_0,M_{1/2},\ldots \sim M_Z$,
and that actually most of the parameter space has already been  excluded~\cite{GRS}.
We are interested in SUSY as long as it solves the hierarchy problem:
therefore we restrict our analysis to sparticle spectra reasonably close
to the weak scale.

As usual this is just a two parameter slicing: 
models have more free parameters (e.g.\ $\tan\beta$, $A$-terms, sign of $\mu$,\ldots)
which are kept fixed at arbitrarily chosen values.
While this is often a critical arbitrary choice,
precision data are again `robust': i.e.\ we will find smooth feature-less contours 
that are rather insensitive to values of the extra parameters.
Therefore we will show a small number of figures.

\begin{figure}
\begin{center}$$\hspace{-5mm}
\includegraphics[width=8.5cm]{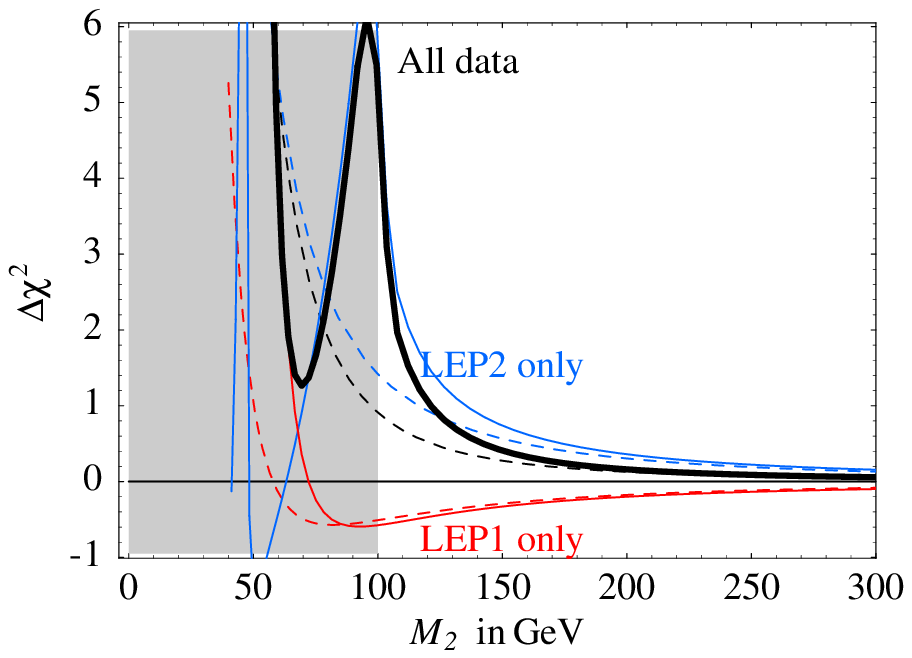}\qquad
\includegraphics[width =8.5cm]{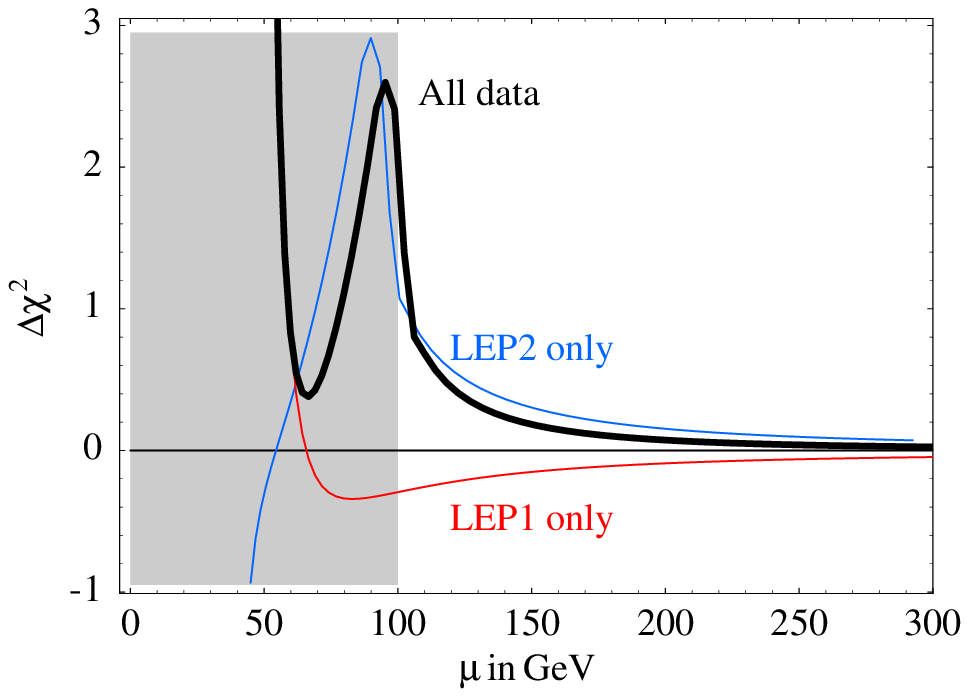}$$
\caption[X]{\em Fig.\fig{SUSYM2}a: $\chi^2-\chi^2_{\rm SM}$ in the case of light gauginos
with $\tan\beta=10$, gaugino unification and $m_h=115\GeV$.  
The dashed line are our analytical approximation,
while the continuous lines are the full numerical computation.
The thick line is the result obtained including all data.
The upper blue line shows the contribution of LEP2 only.
The lower red line shows the result omitting LEP2, including
only `traditional' precision data.
The shaded regions is excluded by direct LEP2 searches.
Fig.\fig{SUSYM2}b: analogous plot for the case of light higgsinos.
We show only the full result.
\label{fig:SUSYM2}}
\end{center}
\end{figure}

\section{`Split' supersymmetry}\label{split}
We start with a simple case: we assume that only fermionic sparticles  are light
so that only corrections to propagators are relevant.
This might be not only a warming exercise:
the  MSSM with heavy scalar sparticles received recent attention~\cite{anthropoids}.
In this limit most MSSM problems get milder, 
most MSSM successes are retained but
SUSY no longer solves the hierarchy `problem'.
This was considered as the most important success of SUSY,
but alternative antrophic interpretations~\cite{Weinberg,Barr} gained credit
in view of recent results:
the possible discovery of a small cosmological constant;
the non-observation of new physics around the Fermi scale;
the realization that string models are even more abundant that what feared.
This anthropic scenario is pudically named `split supersymmetry'.

Although there is no longer a link between the scales of
SUSY breaking and of electroweak symmetry breaking,
we still restrict our attention to fermionic sparticles close to the Fermi scale,
because only in this case precision observables receive detectable corrections.
In the same way, scalar sparticles give negligible effects even if they are 
relatively close to the Fermi scale, so that SUSY can still solve the hierarchy problem.

\medskip

The spectrum  of fermionic sparticles
is specified by $\mu,M_1,M_2,M_3$ and $\tan\beta$.
We assume a GUT relation among gaugino masses, $\tan\beta=10$ and $m_h=115\GeV$.

Let us start from the sub-case in which
only gaugino masses are around $M_Z$ and all other sparticles are much heavier.
In $\hat S,\hat T,W,Y$ approximation we have
\beq\hat{S}=\hat{T}=Y\simeq0,\qquad W \simeq \frac{\alpha_2}{15\pi}\frac{M_W^2}{M_2^2}\eeq
which does not depend on $\tan\beta,M_1,M_3$.
Fitting only traditional precision data (LEP1, SLD, the $W$ mass,\ldots)
gives $W=(0.7\pm0.9)\cdot 10^{-3}$ i.e.\ a
almost $1\sigma$ preference for $M_2\approx 80\GeV$,   
as emphasized in~\cite{MTW} (see also~\cite{Cho}).
Adding LEP2 data this preference disappears because
the best fit shifts towards negative $W$.\footnote{The central value might
shift again when combined $e^+e^-\to e^+e^-$
LEP2 cross section data from all LEP collaborations will be available.}
Going beyond the $\hat S,\hat T,W,Y$ approximation,
this result is confirmed by the exact numerical result, shown in fig.\fig{SUSYM2}a.
We see that in all the experimentally allowed range for the chargino mass, 
$M_\chi \circa{>}100\GeV$,
the $\hat S,\hat T,W,Y$ approximation accurately reproduces the full LEP1 fit.
On the contrary when the lightest chargino or neutralino is slightly above the LEP2 direct limit,
 $M_\chi\approx 100\GeV $, the  $\hat S,\hat T,W,Y$ approximation underestimates 
SUSY corrections to LEP2 observables, because
one loop chargino and neutralino corrections to LEP2 observables
are enhanced by an ${\cal O}(1)$ factor,
by having a virtual chargino or neutralino almost on-shell.
Going to chargino and neutralino masses above the LEP2 direct bound
 the resonant enhancement disappears
and the $\hat S,\hat T,W,Y$ approximation becomes correct.

\bigskip

The same thing happens if only higgsinos are light: in this limit
\beq\hat{S}=\hat{T}\simeq0,\qquad W \simeq Y\simeq \frac{\alpha_2}{30\pi}\frac{M_W^2}{\mu^2}.\eeq
Ignoring LEP2 we agree with~\cite{MTW}; including LEP2 we get the different result
of fig.\fig{SUSYM2}b.

\begin{figure}
\begin{center}LEP2 precision data included\vspace{8mm}
$$\includegraphics[width=8cm]{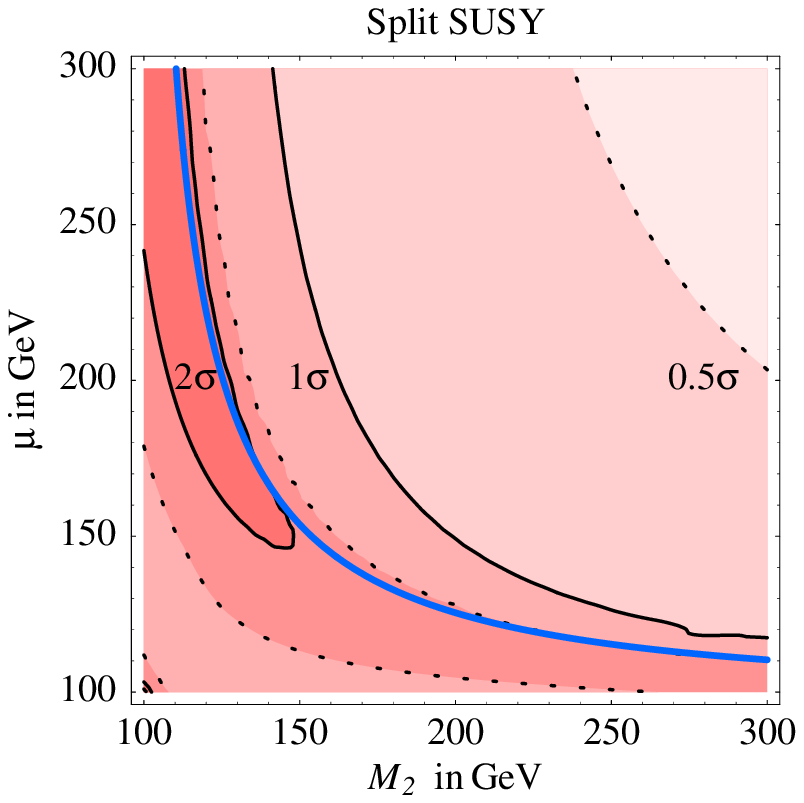}\qquad
\includegraphics[width=8cm]{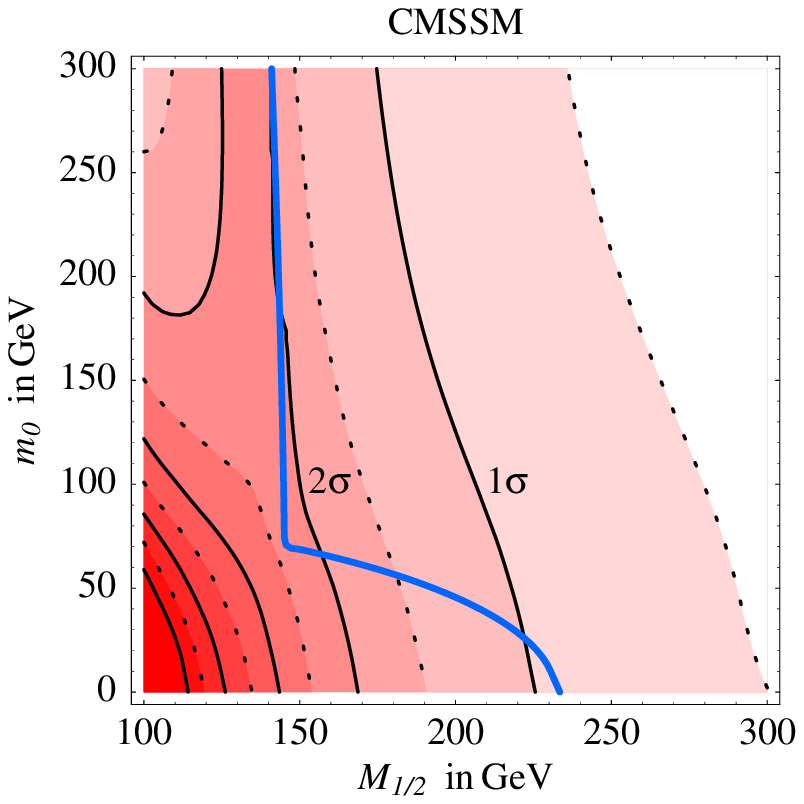}$$
$$\includegraphics[width=8cm]{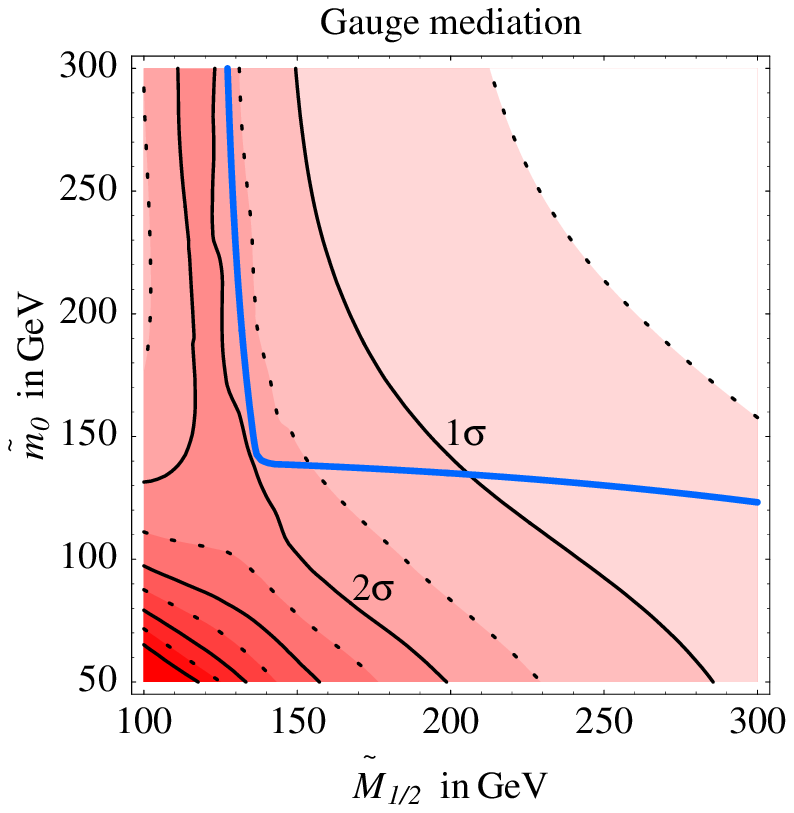}\qquad
\includegraphics[width=8cm]{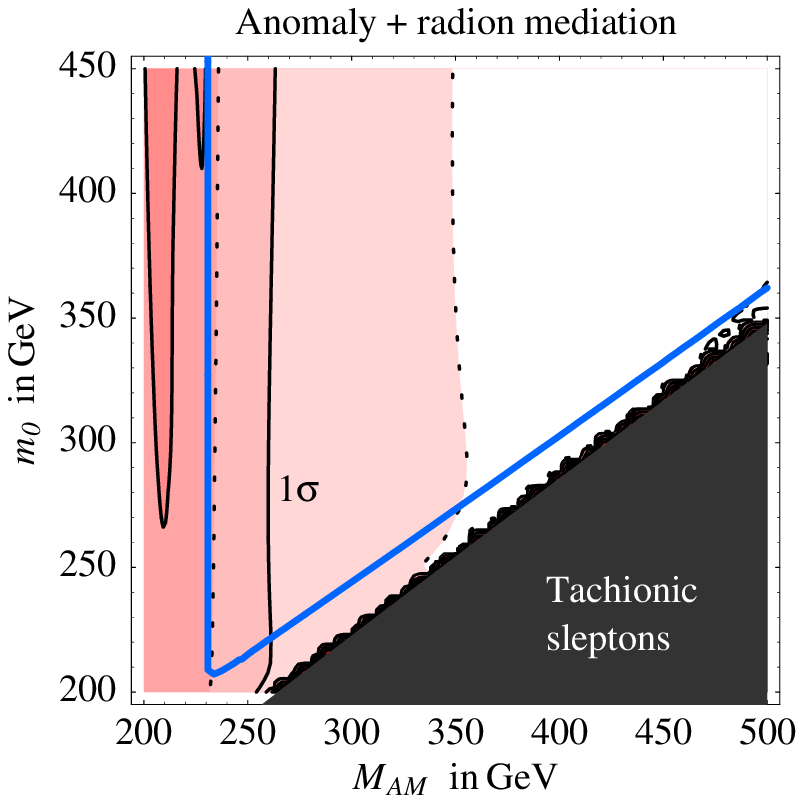}$$
\caption{\em Fits of precision data.
Regions shaded in red are disfavored  at $1,2,3,\ldots\sigma$,
as indicated on the iso-lines.
Regions below the thick blue line are excluded by LEP2 direct searches.
We performed a full one-loop analysis, including LEP2 precision data.
We kept fixed $\tan\beta=10$, $A_0=0$, $\lambda_t(M_{\rm GUT})=0.6$,
$\hbox{\rm sign}\,\mu=+1$,
the gauge-mediation scale $M_{\rm GM} = 10^{10}\GeV$.
\label{fig:SUSY}}
\end{center}
\end{figure}

\begin{figure}
\begin{center}LEP2 precision data not included\vspace{8mm}
$$\includegraphics[width=8cm]{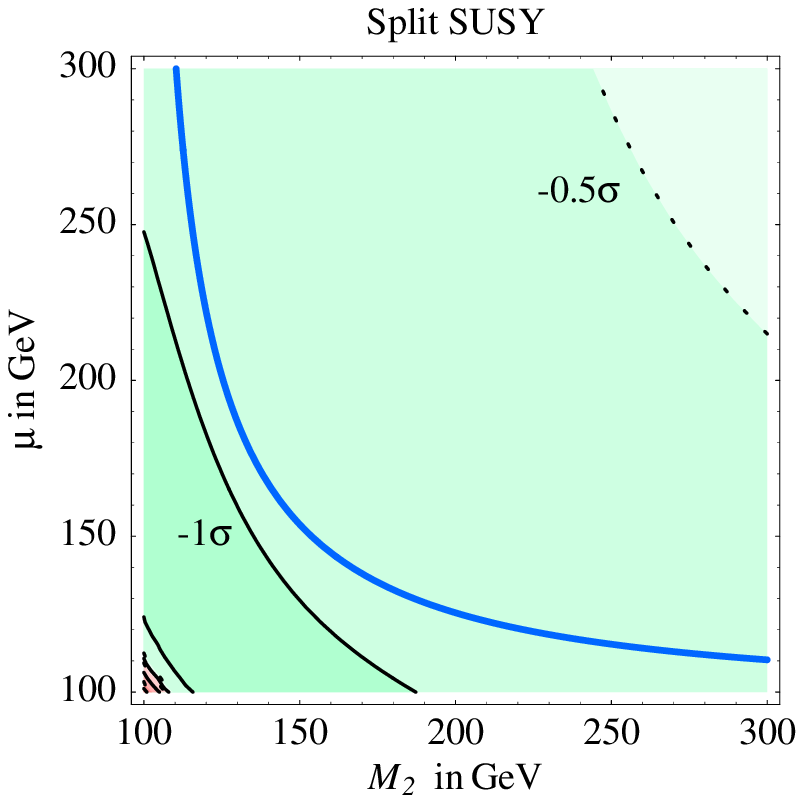}\qquad
\includegraphics[width=8cm]{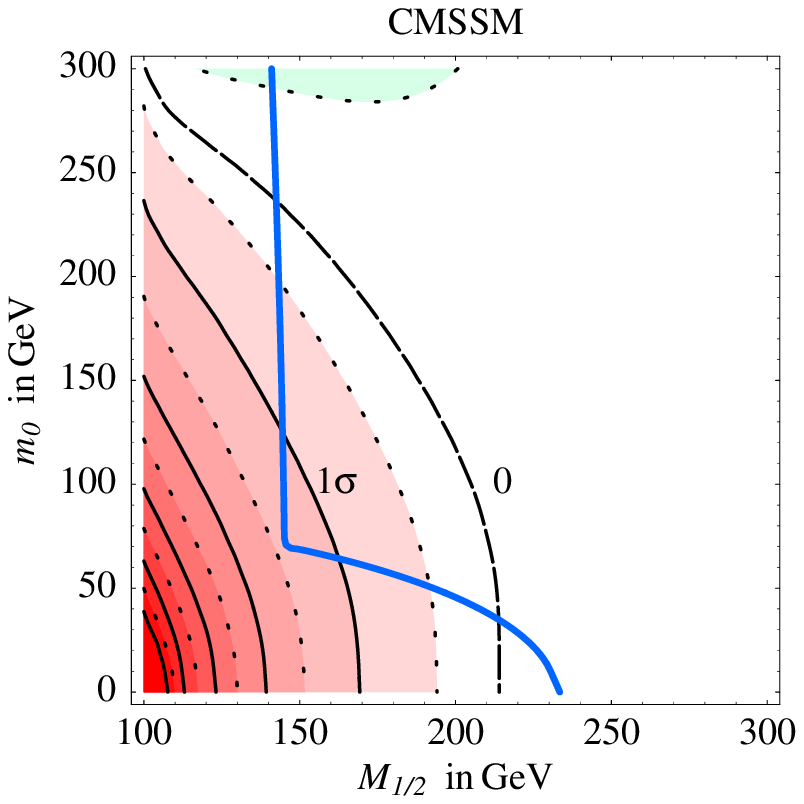}$$
$$\includegraphics[width=8cm]{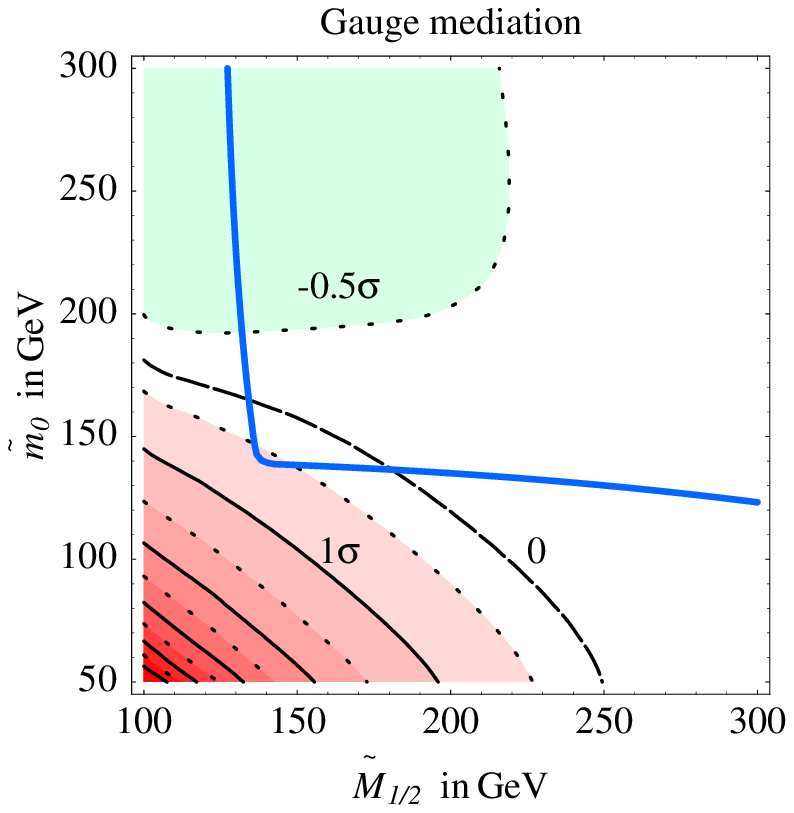}\qquad
\includegraphics[width=8cm]{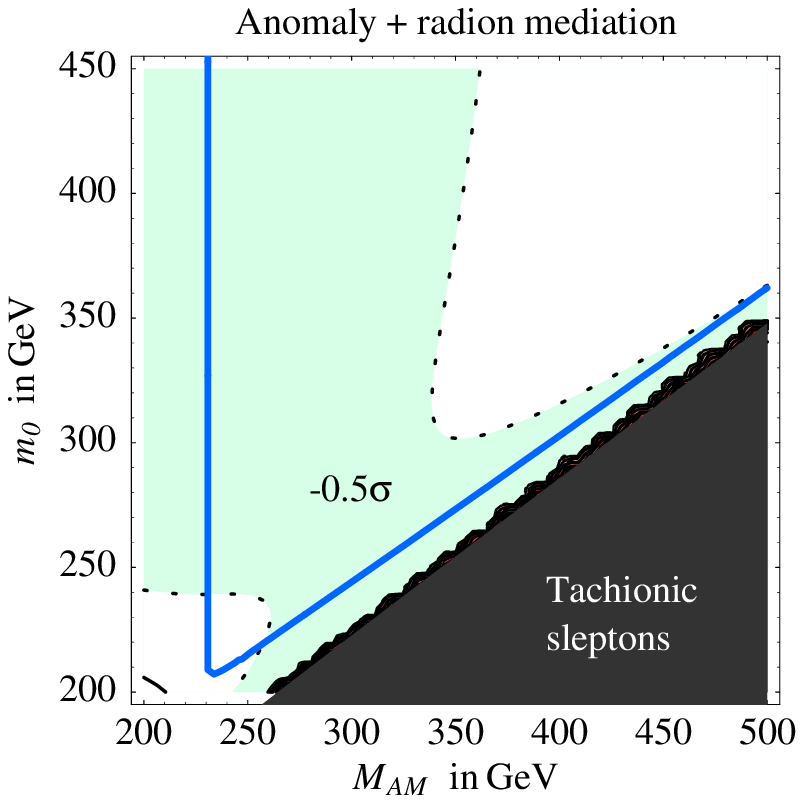}$$
\caption{\em As in fig.\fig{SUSY}, but without including LEP2 precision data.
Regions shaded in green are favored  at $-1,2,3,\ldots\sigma$.
\label{fig:SUSYLEP1}}
\end{center}
\end{figure}

Finally, fig.\fig{SUSY}a shows the global fit of precision data in the ($M_2,\mu$) plane.
We find no favored regions, nor new statistically significant constraints.
Gauginos and higgsinos masses slightly above their bound
from direct searches are mildly disfavored by precision data.
For comparison  fig.\fig{SUSYLEP1}a shows the global fit omitting precision LEP2 data.
Notice that in the `split' SUSY limit there are no corrections to $g_\mu-2$, $b\to s\gamma$,\ldots.

\begin{figure}
\begin{center}
$$\includegraphics[width=8cm]{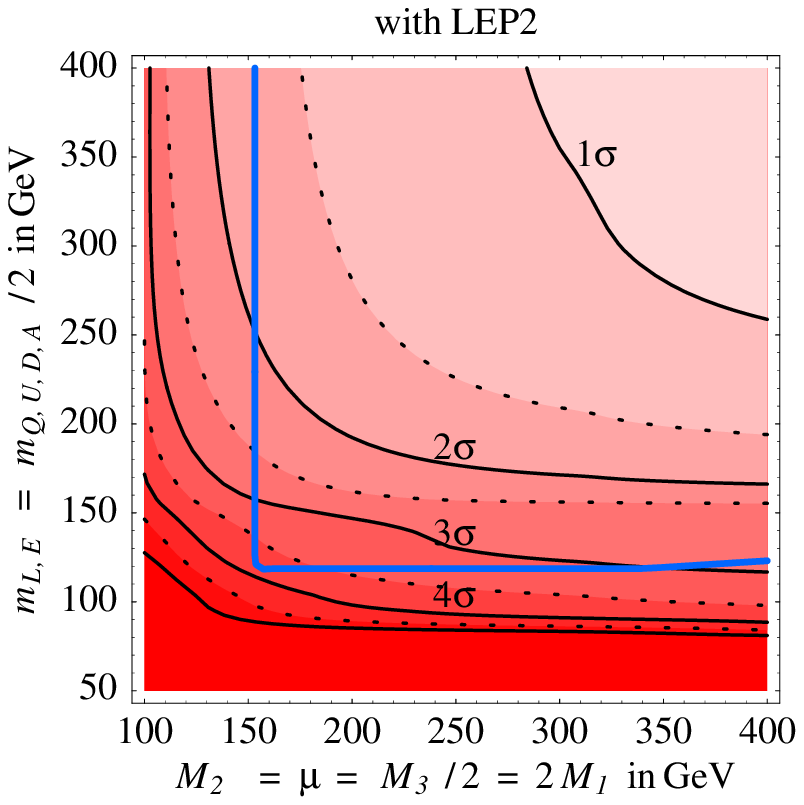}\qquad
\includegraphics[width=8cm]{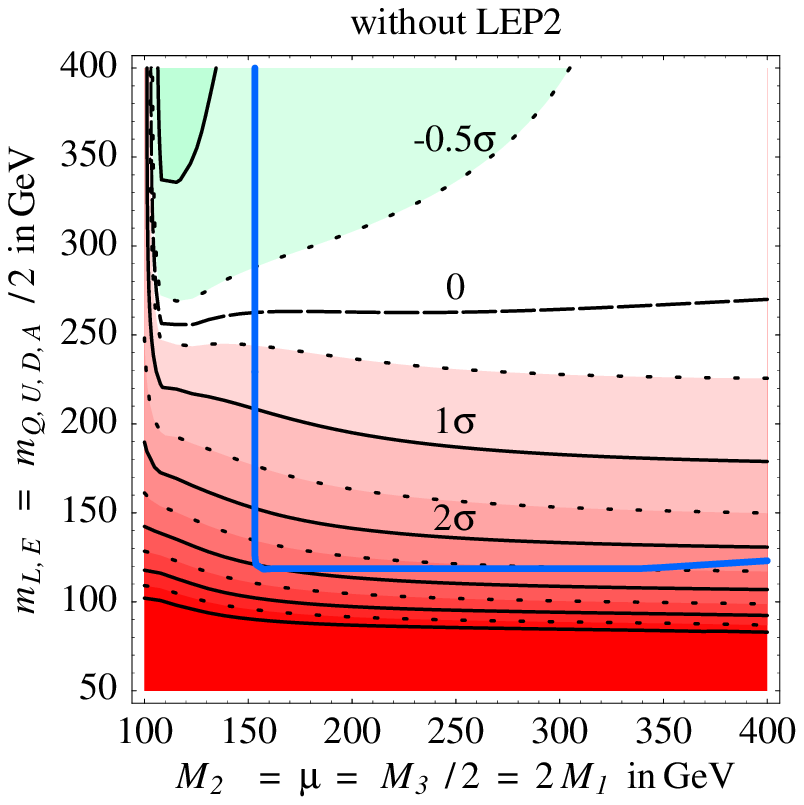}$$
\caption{\em Fit of precision data with the sample sparticle spectrum
of eq.\eq{SUSYLIGHT}.
The label `$3\sigma$' means $\Delta\chi^2 = 3^2$ (i.e.\ SUSY is disfavoured)
and the label `$-0.5\sigma$' means $\Delta\chi^2 = -0.5^2$ (i.e.\ SUSY is favoured).
\label{fig:SUSYLIGHT}}
\end{center}
\end{figure}

\section{A simple sparticle spectrum}\label{LIGHT}
Before considering popular models, we compare data with a 
simple sparticle spectrum, chosen such that
all sparticles can be at the same time close to present direct collider bounds.
We assume
\beq m_E = m_L =m_Q/2 = m_U/2 = m_D/2= m_A/2,\qquad
M_2 = \mu = 2M_1=M_3/2,\qquad \tan\beta=10\label{eq:SUSYLIGHT}\eeq
and vanishing $A$-terms. All parameters are renormalized at the weak scale.
The gluino mass $M_3$ marginally affects precision data,  and could increased
in order to avoid constraints from hadron colliders.
The result is shown in fig.\fig{SUSYLIGHT}a (LEP2 data included)
and in fig.\fig{SUSYLIGHT}b (LEP2 not included).
Here and in the following, we plot iso-lines of
$\Delta\chi^2=\chi^2-\chi^2_{\rm SM}$.
Below the thick line some charged sparticle is lighter than $100\GeV$,
which is excluded by direct searches at LEP2.
We see that precision data disfavor regions where all sparticles
are close to present direct bounds.
In the next sections we analyze popular models, 
finding analogous results.
Constraints will be somewhat weaker because such models
force colored sparticles and Higgsinos to be heavier than what allowed by direct searches.

\section{CMSSM}\label{CMSSM}
Unification of gauge couplings and the non-observation of SUSY flavor effects
suggests the following assumption about the
sparticle spectrum renormalized at the GUT scale $M_{\rm GUT}\approx 2~10^{16}\GeV$:
\beq M_{1,2,3}= M_{1/2},\qquad m_{L,Q,U,D,E,H_{\rm u},H_{\rm d}}^2 = m_0^2,\qquad
A_{\rm e,d,u} =A_0.\eeq
We assume $A_0=0$,
$\tan\beta=10$, $\mu>0$, $\lambda_t(M_{\rm GUT}) = 0.6$ 
(as explained in~\cite{heavym0}).
RGE running generates non vanishing $A$-terms at the weak scale.
Table~\ref{massmodels} lists the values of the low-energy parameters most important  for precision observables.
Fig.\fig{SUSY}b shows the result of the global fit of precision data, including LEP2 precision data.
For comparison,  fig.\fig{SUSYLEP1}b shows the analogous fit omitting LEP2 data:
again there were favored regions with $M_2\sim M_Z$ and/or $|\mu|\sim M_Z$.

We do not show constraints from $b\to s\gamma$, $g_\mu-2$, $B_s\to \mu\bar\mu$,
thermal Dark Matter (DM) abundancy and $m_h$,
which unlike precision data depend significantly on other parameters
($\tan\beta$, $A_0$,\ldots) kept fixed at arbitrary values.
Depending on the precise value of $m_t$,
at multi-TeV $m_0$ there can be a fine-tuned region with the atypical phenomenology of higgsino 
Lightest Supersymmetric Particle (LSP).
We ignore this possibility, as higgsino LSP can be better studied in contexts where $\mu$ is naturally small,
rather than fine-tuning the CMSSM in order to avoid its typical outcome, $M_1<|\mu|$.

\bigskip

We now give an example of the accuracy of the $\hat{S},\hat{T},W,Y$ approximation
by considering  at one specific point the three $\varepsilon_{1,2,3}$ observables.
Since SUSY is not universal, we focus on three `leptonic' observables,
$\varepsilon_{1,2,3}^{\rm lept}$.
These are defined like the usual $\varepsilon_{1,2,3}$~\cite{ABC},
except that $Z$-couplings are extracted only from charged lepton data
(chosen because they are more precisely measured than data about neutrinos and quarks).
This means that we take into account SUSY corrections to $Z,\gamma$ and $W^\pm$ propagators, to $Z$/charged lepton vertices, and to $\mu$-decay.
The specific CMSSM point is chosen such that the thermal LSP abundancy agrees with the DM abundancy~\cite{DM}
(without invoking co-annihilations~\cite{coann}) and such that sparticles are as light as possible.
The point is  $m_0 =100\GeV$, $M_{1/2} = 150\GeV$, $A_0=0$, $\tan\beta=10$.
EW breaking is achieved for $\mu=250\GeV$, and the masses of the
lightest neutralino, chargino and slepton are $58\GeV$, $105\GeV$, $124\GeV$
respectively.
This point is far from all limits in which the  $\hat{S},\hat{T},W,Y$ approximation becomes exact.
The SUSY corrections to the $\varepsilon_{1,2,3}$ observables are
\beq\begin{array}{lccc}
&1000\,\delta\varepsilon_1^{\rm lept} &1000\,\delta\varepsilon_2^{\rm lept} &1000\,\delta\varepsilon_3^{\rm lept}\\
\hbox{$\hat{S},\hat{T},W,Y$ approximation:} &+0.29 & -0.40 & -0.43\phantom{.} \\
\hbox{Full numerical result:} & +0.36 & -0.30 & -0.56.
\end{array}\eeq

\begin{table}$$
\begin{array}{c|ccc}
&\hbox{CMSSM} & \hbox{Gauge mediation at $10^{10}\GeV$} & \hbox{Anomaly + radion mediation}\\ \hline
M_2 & 0.82 M_{1/2} & 0.82 \tilde M_{1/2}^{\phantom{X^X}} &-0.43 M_{\rm AM}\\
m_Q^2 & m_0^2 + 6.2  M_{1/2}^2 & 6.5 \tilde{m}_0^2+5.2 \tilde M_{1/2}^2 & m_0^2 +16 M_{\rm AM}^2\\
m_L^2 & m_0^2+0.52  M_{1/2}^2 & 1.3 \tilde{m}_0^2 +0.24 \tilde M_{1/2}^2 & m_0^2 - 0.37 M_{\rm AM}^2\\
\mu^2 + M_Z^2/2 & 0.17 m_0^2 + 2.6 M_{1/2}^2 & 2.9 \tilde{m}_0^2 +1.7 \tilde M_{1/2}^2 & 0.17 m_0^2 + 10 M_{\rm AM}^2
\end{array}$$
\caption[X]{\em Values of the low-energy parameters most important for determining SUSY corrections to precision data
in the models under study.\label{massmodels}}
\end{table}

\section{Gauge mediation}\label{GM}
Soft terms renormalized around the unknown gauge-mediation scale $M_{\rm GM}$ are predicted as~\cite{GM}
\beq M_i = \frac{\alpha_i}{\alpha}\tilde{M}_{1/2},\qquad
m_R^2 = \frac{ c_R^i M_i^2}{\sqrt{n}}\eeq
where $\tilde M_{1/2}$ is a free parameter,
$i=\{1,2,3\}$,
$c^i_R$ are quadratic Casimir coefficients for the various representations $R$,
$n$ is a free parameter equal to
$n=n_5+3n_{10}$ in models
where a single gauge singlet couples supersymmetry breaking
to $n_5$ copies of messenger fields in the $5\oplus\bar5$
representation of SU(5)
and to $n_{10}$ copies in the $10\oplus\overline{10}$
representation~\cite{GM}.
We re-parameterize it as $1/\sqrt{n} =( \tilde{m}_0/\tilde{M}_{1/2})^2$, such that
$\tilde{m}_0$ is a `scalar mass parameter' analogous to 
 the parameter $m_0$ of CMSSM.
In this way gauge-mediated spectra are described  by two free parameters $(\tilde{m}_0,\tilde{M}_{1/2})$ analogous
to the $(m_0,M_{1/2})$ parameters of the CMSSM.
The resulting spectrum is qualitatively similar to the CMSSM
(for a comparison see e.g.~\cite{GM/SuGra}).

We assume $\tan\beta=10$, $\lambda_t(M_{\rm GUT}) = 0.6$, $\mu>0$ and an intermediate
value of the gauge-mediation scale: $M_{\rm GM} = 10^{10}\GeV$.
Table~\ref{massmodels} list the values of the low-energy parameters most important  for precision observables.
The result of our full analysis is shown in fig.\fig{SUSY}c.
For comparison,  fig.\fig{SUSYLEP1}c shows the analogous fit omitting LEP2 data.

\section{Anomaly and radion mediation}\label{AM}
Generic supergravity mediation of SUSY breaking
does not give the flavour-universal sfermion masses suggested by experimental constraints.
A universal scalar mass $m_0$ can be obtained in particular situations
where some particular supergravity contributions become dominant.
This happens in extra dimensional models with localized fields:
if  MSSM fields are separated from  SUSY-breaking fields by a large enough
extra-dimensional distance 
the unwanted generic supergravity effects are forbidden by locality.
Low energy supergravity gives the dominant effects, which are
flavour universal and computable.
One effect is anomaly mediation~\cite{AM}, which predicts
$$ M_i = b_i g_i^2 M,\qquad m_R^2 = -b_i c_i^R g_i^2 M^2\qquad\hbox{at any scale}$$
(we do not write predictions for $A$-terms nor for sparticles involved in the top Yukawa coupling)
where $b_{1,2,3}=\{33/5,1,-3\}$ are the $\beta$-function coefficients of the MSSM gauge couplings.
Sleptons have negative squared masses so that pure anomaly mediation is ruled out.

\medskip

One-loop supergravity corrections give another effect:
a universal contribution $m_0^2$ to scalar masses.
While in the simplest situations $m_0^2 <0$~\cite{rio},
in some particular cases the radion-mediated contribution
gives a positive $m_0^2$~\cite{RSS}.
Therefore models with anomaly plus radion-mediated SUSY breaking
can give an acceptable sparticle spectrum:
\beq M_i = b_i g_i^2 M_{\rm AM},\qquad m_R^2 = m_0^2-b_i c_i^R g_i^2 M_{\rm AM}^2.\eeq
We assume that this boundary condition holds at the GUT scale and that
$\tan\beta=10$ and $\mu>0$.
Table~\ref{massmodels} list the values of the low-energy parameters most important  for precision observables.
To a good approximation only sleptons and winos are light and can give significant effects.
The result of our full analysis is shown in fig.\fig{SUSY}d.
For comparison,  fig.\fig{SUSYLEP1}d shows the analogous fit omitting LEP2 data.

\section{Conclusions}\label{end}
We studied the corrections to precision data generated by one-loop 
supersymmetric effects.
We performed a full analysis of propagator, vertex and box corrections.
We also developed a simple understanding,
based on the `heavy universal' approximation,
and discussed its accuracy.
In this approximation all SUSY corrections to $Z$-pole, low energy and LEP2 observables
are encoded in four $\hat{S},\hat{T},W,Y$ parameters:
eq.s~(\ref{sys:sfermions},\ref{sys:higgs},\ref{sys:gahi}) give
their explicit analytic  expressions.

Furthermore we added for the first time LEP2 $e\bar{e}\to f\bar{f}$ cross sections to the data-set.
In the global fit LEP2 data have a weight comparable to $Z$-pole
data, already included in previous analyses.
Hints  for supersymmetry emphasized  in some previous analyses~\cite{Cho,A...,MTW}
had a marginal statistical significance, 
and are thereby significantly affected by the inclusion of LEP2 data.
Actually such hints get mostly removed, because SUSY
gives positive corrections to $W,Y$ (i.e.\ reduce LEP2 cross sections),
while LEP2 data favor $W<0$.
Rather than performing a general scan searching for more
 fluctuating hints, we preferred to analyze specific models: 
`split' SUSY, the CMSSM, gauge mediation and anomaly plus radion mediation.

The analytic approximation shows that SUSY corrections to precision observables
mainly depend on a few SUSY parameters: left-handed slepton and squark masses, $M_2$ and $\mu$.
(A large stop mixing and a small $\tan\beta$ would also play a significant r\^ole).
No big enhancements or suppressions are possible,
unlike in the cases of other indirect tests
($g_\mu-2$, $b\to s\gamma$, $B_s\to\mu\bar\mu$, dark-matter abundancy, \ldots).
Therefore by varying  2 main parameters
(such as $m_0$ and $M_{1/2}$) keeping fixed all other parameters
we produced plots which represent the general situation in a given model,
rather than being only sample slices of a vast parameter space.
For the same reasons such plots do not show peculiar features.
In `split' SUSY  the only light sparticles are gauginos  and higgsinos .
In anomaly mediation the lightest sparticles
that give the main corrections to precision observables are gauginos and sleptons.
In CMSSM and gauge mediation also stops play a significant r\^ole.
The relevance of LEP2 is clearly seen by comparing our full results of fig.\fig{SUSY}
with fig.\fig{SUSYLEP1}, where LEP2 is not included.
We also analyzed a model where all sparticles can be as light as allowed by direct constraints:
fig.\fig{SUSYLIGHT} shows that this case is disfavored by precision data.

\paragraph{Acknowledgments}
We thank G.\ Altarelli, P.\ Gambino, G. Giudice and R. Rattazzi for useful discussions.
The research of G.M.\ is 
supported by the USA Department of Energy, under contract W-7405-ENG-36.
A.S.\ thanks T.\ Hahn for help in installing the LoopTools code~\cite{LoopTools}.

\appendix

 \section{Technical details}\label{tech}
 One-loop computations of LEP2 cross sections
 have been performed using the  FeynArts and FormCalc  codes~\cite{FormCalc}.
 Loop functions have been computed using the LoopTools code~\cite{LoopTools}.
The set of data we fit is basically the same as in~\cite{Barbieri:2004qk}.
We compute LEP2 cross sections at the various energies around $200\GeV$
corresponding to the LEP2 runs.

 Our full analysis of LEP1 data is performed following the lines of~\cite{ABC},
 while  our full analysis of LEP2 data is performed following the lines of~\cite{MSSMLEP2}.
 Unfortunately these two strategies employ  different renormalization procedures,
 so that care is needed to combine the two analyses.
 The basic difference is that~\cite{ABC} fixes the three basic SM parameters
 $v,g,g'$ using the three observables $\alpha,M_Z,G_{\rm F}$,
 while~\cite{MSSMLEP2} employs the alternative set $\alpha,M_Z,M_W$.
 As described in~\cite{Denner} the second procedure is implemented by subtracting
gauge boson propagators as
\begin{eqnsystem}{sys:Denner}
\Pi_{WW}^{\rm ren}(p^2) &=& \Pi_{WW}(p^2) -\Pi_{WW}(M_W^2) - (p^2-M_W^2)\Pi'(M_W^2)\\
\Pi_{ZZ}^{\rm ren}(p^2) &=& \Pi_{ZZ}(p^2) -\Pi_{ZZ}(M_Z^2) - (p^2-M_Z^2)\Pi'(M_Z^2)\\
\Pi_{AA}^{\rm ren}(p^2) &=& \Pi_{AA}(p^2) -\Pi_{AA}(0) - p^2 \Pi'_{AA}(0)\\
\Pi_{AZ}^{\rm ren}(p^2) &=& \Pi_{AZ}(p^2) -\Pi_{AZ}(0) - p^2 [ \Pi_{AZ}(M_Z^2)-\Pi_{AZ}(0)]/M_Z^2
\end{eqnsystem}
such that $\alpha,M_Z,M_W$ keep their tree level values.

In order to perform a global fit one has to match the two different strategies.
We convert LEP2 observables, computed using the $\alpha,M_Z,M_W$ scheme~\cite{MSSMLEP2},
 to the $\alpha,M_Z,G_{\rm F}$ scheme.
Since LEP2 $e\bar{e}\to f\bar{f}$ cross sections do not directly depend on $M_W$,
the only difference amounts to a shift in the values of the $Z$-couplings,
as predicted in terms of the chosen set of basic observables.
In both schemes $Z$-couplings are conveniently parameterized by an auxiliary
effective weak angle, defined as
\beq \sW(\alpha,M_Z,G_{\rm F}) \equiv \frac{1}{2}\left[1-\sqrt{1-\frac{4\pi\alpha}{\sqrt{2} G_F M_Z^2}}\right],
\qquad
\sW (\alpha,M_Z,M_W) \equiv \sqrt{1- \frac{M_W^2}{M_Z^2}}.\eeq
At tree level these two definitions are equivalent, while at one-loop order they differ by
\beq \delta \sW^2\equiv 
\sW^2 (\alpha,M_Z,M_W) -\sW^2(\alpha,M_Z,G_{\rm F}) = 
2\sW^2 \cW^2 \bigg[\frac{\varepsilon_3-\varepsilon_1 \cW^2/2\sW^2} {\cW^2-\sW^2}+ \frac{\varepsilon_2}{2\sW^2}\bigg].\eeq
One-loop predictions for LEP2 observables $\sigma$ are then converted as
\beq\sigma(\alpha,M_Z,G_{\rm F})=\sigma(\alpha,M_Z,M_W) -
\frac{\partial\sigma}{\partial\sW^2}\delta \sW^2.\eeq

\frenchspacing\footnotesize\begin{multicols}{2}
\end{multicols}
\end{document}